\documentclass[12pt]{article}
\usepackage{amsmath,amssymb}
\usepackage{graphicx}
\usepackage[pdftex]{hyperref}
\DeclareGraphicsExtensions{.eps,.bmp,.wmf,.jpg,.pdf}
\numberwithin{equation}{section}
\def\be{\begin{equation}}
\def\ee{\end{equation}}

\def\bea{\begin{eqnarray}}
\def\eea{\end{eqnarray}}

\title{New infrared cut-off for the holographic scalar fields models of dark energy}
\author{L.N. Granda\thanks{ngranda@univalle.edu.co} \, and\  A. Oliveros\thanks{alexogar@univalle.edu.co}\\
Department of Physics, Universidad del Valle\\ A.A. 25360, Cali,
Colombia} 
\date{}
\begin{document}
\maketitle

\begin{abstract}
\noindent Introducing a new infrared cut-off for the holographic dark-energy, we study the correspondence between the quintessence, tachyon, K-essence and dilaton energy density with this holographic dark energy density in the flat FRW universe. This correspondence allows to reconstruct the potentials and the dynamics for the scalar fields models, which describe accelerated expansion.
\end{abstract}

\section{Introduction}
\noindent 
Recent astrophysical data from distant Ia supernovae, Large Scale Structure and Cosmic Microwave Background observations \cite{SN},\cite{riess},\cite{spergel}, \cite{tegmark}, \cite{abazajian} show that the current Universe is not only expanding, it is accelerating due to some kind of  negative-pressure form of matter known as dark energy (\cite{copeland},\cite{turner}). The simplest candidate for dark energy is the cosmological constant \cite{weinberg}, conventionally associated with the energy of the vacuum with constant energy density and pressure, and an equation of state $w=-1$. The present observational data favor an equation of state for the dark energy with parameter very close to that of the cosmological constant. The next simple model proposed for dark energy is the quintessence (see \cite{copeland1}, \cite{caldwell}, \cite{zlatev}), a dynamical scalar field which slowly rolls down in a flat enough potential. The equation of state for a spatially homogeneous quintessence scalar field satisfies $w>-1$ and therefore can produce accelerated expansion. This field is taken to be extremely light which is compatible with its homogeneity and avoids the problem with the initial conditions. Other scalar field models proposed to explain the nature of the dark energy, are related with K-essence models based on scalar field with non-standard kinetic term \cite{armendariz},\cite{chiba}; string theory fundamental scalars known as tachyon \cite{padmana} and dilaton \cite{gasperini}; scalar field with negative kinetic energy, which provides a solution known as phantom dark energy \cite{caldwell1}. Another class
of dark energy models involve non-standard equations of state \cite{kamen},\cite{odin} (for a review on above mentioned and other approaches, see \cite{copeland}). In all this models of scalar fields the cosmological dynamics is defined once the potential is proposed. 

Another alternative to the solution of the dark energy problem, is related with some facts of the quantum gravity theory known as the holographic principle (\cite{beckenstein, thooft, bousso, cohen, susskind}). This principle emerges as a new paradigm in quantum gravity and was first put forward by t' Hooft \cite{thooft} in the context of black hole physics and later extended by Susskind \cite{susskind} to string theory. According to the holographic principle, the entropy of a system scales not with it's volume, but with it's surface area. In the cosmological context,
the holographic principle will set an upper bound on the entropy of the universe \cite{fischler}. In the work \cite{cohen}, it was suggested that in quantum field theory a short distance cut-off is related to a long distance cut-off (infra-red cut-off $L$) due to the limit set by formation of a black hole, namely, if is the quantum zero-point energy density caused by a short distance cut-off, the total energy in a region of size $L$ should not exceed the mass of a black hole of the same size, thus $L^3\rho_\Lambda\leq LM_p^2$. Thus, if we take the whole universe into account, then the vacuum energy related to this holographic principle is viewed as dark energy, usually called holographic dark energy. The largest $L$ allowed is the one saturating this inequality so that we get the holographic dark energy density
\begin{equation}\label{eq1}
\rho_\Lambda=3c^2M_p^2L^{-2}
\end{equation}
where $c^2$ is a numerical constant and $M_p^{-2}=8\pi G$.
In this paper we use a new IR cut-off proposed in \cite{granda} for the holographic dark energy
\begin{equation}\label{eq2}
\rho_{\Lambda}=3M_p^2\left(\alpha H^2+\beta \dot{H}\right)
\end{equation}
where $H=\dot{a}/a$ is the Hubble parameter and $\alpha$ and $\beta$ are constants which must satisfy the restrictions imposed by the current observational data. Besides the fact that the underlying origin of the holographic dark energy is still unknown, the inclusion of the time derivative of the Hubble parameter may be expected as this term appears in the curvature scalar (see \cite{gao}), and has the correct dimension. This kind of density may appear as the simplest case of more general $f(H,\dot{H})$ holographic density in the FRW background. By other hand, contrary to the IR cut-off given by the event horizon \cite{li}, this model avoids the causality problem. The coincidence problem may also be solved as will be clear from the behavior of the scale parameter $a$.
In this paper, using the correspondence with the holographic dark energy density we find the explicit form of the fields and potentials for the quintessence, tachyon, K-essence and dilaton fields, which determine the cosmological dynamics of this models. 

\section{The model}
Let us start with the Friedman equation taking into account the holographic dark energy density given by (\ref{eq2}) \cite{granda}. Restricting our study to the current cosmological epoch, in this work we are not considering the contributions from matter and radiation and we assumed that the dark energy $\rho_{\Lambda}$ dominates, thus the Friedman equation becomes simpler. From our expression for the holographic dark energy (\ref{eq2}) it follows the Friedman equation
\begin{equation}\label{eq3}
H^2=\alpha H^2+\beta \dot{H}
\end{equation}
and integrating this equation with respect to the cosmological time $t$ we obtain
\begin{equation}\label{eq4}
H=\frac{\beta}{\alpha-1}\frac{1}{t}
\end{equation}
which gives rise to the power-law expansion $a\propto t^{\beta/(\alpha-1)}$. Similar expression for $H$ was obtained in \cite{li} using the future event horizon as the infrared cut-off, and in \cite{nojiri} in the small $t$ limit of an infrared cut-off that depends on local and non-local quantities. By other hand, from the conservation equation 
\begin{equation}
\frac{\partial\rho_{\Lambda}}{\partial t}+3H(\rho_{\Lambda}+p_{\Lambda})=0
\end{equation}
and using a barotropic equation of state for the holographic energy and pressure densities $p_{\Lambda}=\omega_{\Lambda}\rho_{\Lambda}$, 
we obtain an expression for the EoS parameter $\omega_{\Lambda}$
\begin{equation}\label{eq5}
\omega_{\Lambda}=-1-\frac{2\alpha H\dot{H}+\beta\ddot{H}}{3H\left(\alpha H^2+\beta\dot{H}\right)}
\end{equation}
replacing $H$ from Eq. (\ref{eq4}) in (\ref{eq5}) one can write
\begin{equation}\label{eq6}
\omega_{\Lambda}=-1+\frac{2}{3} \frac{\alpha-1}{\beta}
\end{equation}
which express $\omega_{\Lambda}$ in terms of the constants $\alpha$ and $\beta$. In order to obtain accelerated expansion, the constants $\alpha$ and $\beta$ must satisfy the restrictions followed from the Eq. (\ref{eq4},\ref{eq6}): $\beta>\alpha-1$ if $\alpha>1$ or $\beta<\alpha-1$ if $\alpha<1$, if we consider the $-1<\omega_{\Lambda}<-1/3$ phase. Note that for $\alpha<1$, $\beta>0$ or $\alpha>1$, $\beta<0$ the holographic density describes a phantom-like phase of the evolution with $w_{\Lambda}< -1$.

\section{Correspondence with scalar field models}
In  this section we establish a correspondence between our proposal for the holographic density and various scalar field models, by comparing the holographic density with the corresponding scalar field model density and also equating the equations of state for this models with the EoS parameter given by (\ref{eq6}). In this work we are not considering the contributions from matter and radiation to the Friedman equation, and for this reason the solution presented here differs from the one presented in the work \cite{granda}. Therefore we will not fix the constants here and  will impose conditions on $\alpha$ and $\beta$ dictated by the existence in  each model of attractor solutions giving accelerated expansion.
\subsection*{Holographic quintessence model}
In the flat Friedman background the energy density and pressure density of the scalar field are given by \cite{copeland}
\begin{equation}\label{eq7}
\rho=\frac{1}{2}\dot{\phi}^2+V(\phi)    \ \ \,\ \ \  p=\frac{1}{2}\dot{\phi}^2-V(\phi)
\end{equation}
The equation of state parameter for the scalar field is given by
\begin{equation}\label{eq8}
\omega_{\phi}=\frac{\dot{\phi}^2-2V(\phi)}{\dot{\phi}^2+2V(\phi)}
\end{equation}
which compared with the holographic EoS parameter (\ref{eq6}) gives the equation
\begin{equation}\label{eq9}
\frac{\dot{\phi}^2-2V(\phi)}{\dot{\phi}^2+2V(\phi)}=-1+\frac{2}{3} \frac{\alpha-1}{\beta}
\end{equation}
which together with the equation
\begin{equation}\label{eq10}
\rho_{\phi}=\frac{1}{2}\dot{\phi}^2+V(\phi)=3M_p^2\left(\alpha H^2+\beta \dot{H}\right)
\end{equation}

can be solved to obtain the explicit expressions for the scalar field and the potential, namely
\begin{equation}\label{eq11}
\phi=\left(\frac{2\beta}{\alpha-1}\right)^{1/2}M_p \ln t
\end{equation}
and 
\begin{equation}\label{eq12}
V(\phi)=\frac{3\beta-\alpha+1}{(\alpha-1)^2}M_p^2\exp\left(-\sqrt{\frac{2(\alpha-1)}{\beta}}\frac{\phi}{M_p}\right)
\end{equation}
A detailed analysis of the cosmological dynamics of an exponential potential is given in \cite{copeland}. This potential can produce an accelerated expansion provided that $\beta/(\alpha-1)>1$ (see Eq. (\ref{eq9})) and also has cosmological scaling solutions \cite{copeland1}. In the context of the phase-space analysis as presented in (\cite{copeland}), the exponential potential for the scalar field has attractor solutions which give rise to an accelerated expansion if $\alpha$ and $\beta$ satisfy $(\alpha-1)/\beta <1$ which is the same condition required by the power-law accelerated expansion.

\subsection*{Holographic tachyon model}
The tachyon field has been proposed as the
source of dark energy \cite{padmana1}, \cite{abramo} and may be described by effective field theory corresponding to some sort of tachyon condensate with an effective Lagrangian density given by \cite{roo}, \cite{sen}.
\begin{equation}\label{eq13}
{\cal L}=-V(\phi)\sqrt{1+\partial_{\mu}\phi\partial^{\mu}\phi}
\end{equation}
In a flat FRW background the energy density $\rho$ and the pressure density $p$ are given by
\begin{equation}\label{eq14}
\rho=\frac{V(\phi)}{\sqrt{1-\dot{\phi}^2}}  \ \ \,\ \ \   p=-V(\phi)\sqrt{1-\dot{\phi}^{2}}
\end{equation}
where $V(\phi)$ is the tachyon potential. From Eqs. (\ref{eq14}) follows the tachyon equation of state parameter 
\begin{equation}\label{eq15}
\omega_T=\dot{\phi}^2-1
\end{equation}
Establishing the correspondence between the holographic dark energy and tachyon energy density, using the Eqs. (\ref{eq2},\ref{eq14}) for the respective densities
\begin{equation}\label{eq16}
\rho_{\Lambda}=3M_p^2\left(\alpha H^2+\beta \dot{H}\right)=\frac{V(\phi)}{\sqrt{1-\dot{\phi}^2}}
\end{equation}
and comparing the EoS parameters Eqs. (\ref{eq6}, \ref{eq15}) $\dot{\phi}^2-1=\omega_{\Lambda}$, one obtains after integration with respect to $t$
\begin{equation}\label{eq17}
\phi=\sqrt{\frac{2(\alpha-1)}{3\beta}}t
\end{equation}
where we assumed the integration constant equal to zero.
Therefore, using Eq. (\ref{eq16}) and the expression (\ref{eq4}) for $H$ we obtain for the tachyon potential in terms
of the scalar field (\ref{eq17})
\begin{equation}\label{eq18}
V(\phi)=2M_p^2\left(1-\frac{2(\alpha-1)}{3\beta}\right)^{1/2}\left(\frac{\beta}{\alpha-1}\right)\frac{1}{\phi^2}
\end{equation}
this inverse square potential also corresponds to the potential obtained for scaling solutions in the context of brane world cosmology \cite{sami}, \cite{copeland2}. Considering the phase-space analysis, note that for the tachyon system with the inverse square potential in the case of a scalar-field dominated solution (see \cite{copeland},\cite{copeland3}), the condition for accelerated expansion translates into the condition 
\begin{equation}\label{eq19}
2\frac{\alpha-1}{\beta}\left[1-2\frac{\alpha-1}{\beta}\right]^{-1/2}<\frac{2}{3}
\end{equation}
or equivalently $-1/9(1+\sqrt{10})<(\alpha-1)/\beta<1/9(-1+\sqrt{10})$, which gives the only viable late-time attractor solution (\cite{copeland},\cite{naskar}). Note that this restriction is consistent with the one imposed by the power-law accelerated expansion $(\alpha-1)\beta<1$ and with $(\alpha-1)\beta<1/2$, so that the square root is well defined . An holographic correspondence with the tachyon, phantom and Chaplygin gas models using the event horizon, was poposed in \cite{setare},\cite{setare1},\cite{setare2}.

\subsection*{Holographic $K$-essence model}
The scalar field model known as k-essence is also used to explain the observed late-time acceleration of the universe. It is well known that k-essence scenarios have attractor-like dynamics, and therefore avoid the fine tuning of the initial conditions for the scalar field (\cite{mukhanov}).
This kind of models is characterized by non-standard kinetic energy terms, and are described by a general scalar field action which is a function of $\phi$ and $X=-1/2\partial_{\mu}\phi\partial^{\mu}\phi$, and is given by \cite{damour}
\begin{equation}\label{eq20}
S=\int d^4x\sqrt{-g}p(\phi,X)
\end{equation}
where $p(\phi,X)$ corresponds to a pressure density and usually is restricted to the Lagrangian density of the form
$p(\phi,X)=f(\phi)g(X)$. Based on the analysis of the low-energy effective action of string theory (see \cite{damour} for details) the
Lagrangian density can  be transformed into
\begin{equation}\label{eq21}
p(\phi,X)=f(\phi)\left(-X+X^2\right)
\end{equation}
From the energy momentum-tensor for this Lagrangian density, follows the next expression for the energy density
of the field $\phi$ (see \cite{damour})
\begin{equation}\label{eq22}
\rho(\phi,X)=f(\phi)\left(-X+3X^2\right)
\end{equation}
And the equation of state using (\ref{eq21}) and (\ref{eq22}) is given by
\begin{equation}\label{eq23}
\omega_K=\frac{X-1}{3X-1}
\end{equation}
Equating this parameter with the holographic EoS parameter (\ref{eq6}) $\omega_K=\omega_{\Lambda}$ we find the solution for $X$
\begin{equation}\label{eq24}
X=\frac{1}{3} \left(\frac{3\beta-\alpha+1}{2\beta-\alpha+1}\right)
\end{equation}
which shows that $X$ is constant. The condition $X<2/3$ which gives rise to an accelerated expansion, translates into $\alpha-1<\beta$. The equation (\ref{eq24}) can be solved to obtain the expression for the scalar field in the flat FRW background
\begin{equation}\label{eq25}
\phi=\left[\frac{2}{3}\left(\frac{3\beta-\alpha+1}{2\beta-\alpha+1}\right)\right]^{1/2}t
\end{equation}
where we have taken the integration constant $\phi_0$ equal to zero.\\
\noindent Using the correspondence between K-essence and holographic energy densities, Eqs (\ref{eq22}) and (\ref{eq2}) ($\rho_{\Lambda}=\rho(\phi,X)$) and replacing $X$ by Eq. (\ref{eq24}) and $H$ from (\ref{eq4}) we get a simple equation for $f(\phi)$ with solution given by
\begin{equation}\label{eq26}
f(\phi)=6M_p^2\beta\left[\frac{2\beta-\alpha+1}{(\alpha-1)^2}\right]\frac{1}{\phi^2}
\end{equation}
where we used the Eq. (\ref{eq25}) for $\phi$. 
Hence, as a result of our holographic-K-essence correspondence one obtains the k-essence potential $f(\phi)$ given by (\ref{eq26}), which is a result of the power-law expansion.

\subsection*{Holographic dilaton field}
This model is described by the pressure (Lagrangian) density
\begin{equation}\label{eq27}
p_D=-X+ce^{\lambda\phi}X^{2}
\end{equation}
where $c$ is a positive constant and $X=1/2\dot{\phi}^2$. This model appears from a four-dimensional effective low-energy string action \cite{piazza}
and includes higher-order kinetic corrections to the tree-level action in low energy effective string theory. The correspondence between  the dilaton energy density given by $\rho_D=-X+3ce^{\lambda\phi}X^{2}$ (see \cite{piazza}), and the holographic energy density \ref{eq2} gives the equation
\begin{equation}\label{eq28}
\rho_D=-X+3ce^{\lambda\phi}X^{2}=3M_p^2\left(\alpha H^2+\beta \dot{H}\right)
\end{equation}
and the correspondence with the holographic dark energy equation of state is written as
\begin{equation}\label{eq29}
\omega_D=\frac{-1+ce^{\lambda\phi}X}{-1+3ce^{\lambda\phi}X}=-1+\frac{2}{3}\frac{\alpha-1}{\beta}
\end{equation}
Solving this equation with respect to $\phi$ and integrating with respect to $t$ we find
\begin{equation}\label{eq30}
\phi=\frac{2}{\lambda}\ln \left[\frac{\lambda}{\sqrt{6c}}\left(\frac{3\beta-\alpha+1}{2\beta-\alpha+1}\right)^{1/2}t\right]
\end{equation}
replacing $X e^{\lambda\phi}$ from Eq. (\ref{eq29}) and using Eqs. (\ref{eq30}) and (\ref{eq4}) in the Eq. (\ref{eq28}), we can express $\lambda M_p$ in terms of $\alpha$ and $\beta$.
\begin{equation}\label{eq31}
\lambda M_p=\sqrt{\frac{2}{3}}\left[\frac{1}{\beta(2\beta-\alpha+1)}\right]^{1/2}(\alpha-1)
\end{equation}
which is well defined in the region of interest $\beta>(\alpha-1)$ (with $\beta>0$, so that the square root is well defined).
In the dynamics of the autonomous system for this model, the condition for accelerated expansion corresponds to $\lambda M_p<\sqrt{6}/3$ (see \cite{copeland}), which translates into the condition
$-2<(\alpha-1)/\beta<1$. In the case $\lambda=0$, which corresponds to $\alpha=1$, the original ghost condensate scenario with $p=-X+X^2$ is recovered, and $\omega_{\Lambda}=\omega_D=-1$.

\section{Discussion}

We propose an infrared cut-off for the holographic dark energy model, which includes a term proportional to $\dot{H}$. Contrary to the holographic dark energy based on the event horizon, this model depends on local quantities, avoiding in this way the causality problem. In the case of dark energy dominance, the power-law expansion appears as the solution to the Friedman equations and this avoids conflict with the coincidence problem. Our proposal automatically generates scalar-field potentials which give rise to scaling solutions in a FRW cosmological background. We found in the case of quintessence that the potential has an exponential form which has attractor solutions giving rise to accelerated expansion for $\beta/(\alpha-1)>1$; for the tachyon field, the potential has an inverse squared form, which contains a late-time attractor solution in the region of the constants, satisfying \ref{eq19}. The potential for the k-essence holographic correspondence has also an inverse squared dependence on the field given by \ref{eq26}, and the model \ref{eq21} with this potential, has an attractor solution with accelerated expansion for $\beta/(\alpha-1)>1$ (see \cite{damour}). We also considered the dilaton condensate without potential, and used the correspondence to find the form of the scalar field and established the region for the constants in which we can expect scaling solutions giving rise to accelerated expansion. In conclusion, with the help of this proposal we have reconstructed the potentials for some scalar field models of dark energy and all this  favors the proposed infrared cut-off as a viable phenomenological model of holographic density.

\section*{Acknowledgments}
This work was supported by the Universidad del Valle, under project CI-7713

\end{document}